\documentclass[journal]{IEEEtran}
\usepackage{upgreek}
\usepackage{graphicx}
\usepackage{subfigure}
\usepackage{amsmath,bm}
\usepackage{amssymb}

\usepackage{enumerate}

\usepackage[square, comma, sort&compress, numbers]{natbib}

\usepackage{color}

\begin{document}

\title{\huge Retro-Reflective Beam Communications with\\ Spatially Separated Laser Resonator}
\author{\normalsize Mingliang Xiong, Mingqing Liu, Qingwei Jiang, Jie Zhou, Qingwen Liu*,~\IEEEmembership{Senior Member,~IEEE}, and  Hao Deng


\thanks{
	* The corresponding author: Qingwen Liu.
}
\thanks{
	M. Xiong, M. Liu, Q. Jiang, J. Zhou, and Q. Liu
	are with the College of Electronics and Information Engineering, Tongji University, Shanghai 201804, China
	(xiongml@tongji.edu.cn; clare@tongji.edu.cn; jiangqw@tongji.edu.cn; jzhou@tongji.edu.cn; and   qliu@tongji.edu.cn). H. Deng is with the School of Software Engineering, Tongji University, Shanghai 201804, China (denghao1984@tongji.edu.cn).}

\thanks{This work was supported by the National Key Research and Development	Project under Grant 2020YFB2103900 and Grant 2020YFB2103902. It was also supported by the National Natural Science Foundation of China under Grant 62071334 and Grant 61771344.}
}

\maketitle

\begin{abstract}
Optical wireless communications (OWC) utilizing infrared or visible light as the carrier attracts great attention in 6G research. Resonant beam communications (RBCom) is an OWC technology which simultaneously satisfies the needs of non-mechanical mobility and high signal-to-noise ratio~(SNR). It has the self-alignment feature and therefore avoids positioning and pointing operations. However, RBCom undergoes echo interference. Here we propose an echo-interference-free RBCom system design based on second harmonic generation. The transmitter and the receiver constitute a spatially separated laser resonator, in which the retro-reflective resonant beam is formed and tracks the receiver automatically. This structure provides the channel with adaptive capability in beamforming and alignment, which is similar to the concept of intelligent reflecting surface  (IRS) enhanced communications, but without hardware and software controllers. Besides, we establish an analytical model to evaluate the beam radius, the beam power, and the channel capacity. The results show that our system achieves longer distance and smaller beam diameter for the transmission beyond $10$ Gbit/s, compared with the existing OWC technologies.
\end{abstract}


\section{Introduction}\label{sec:intro}

\IEEEPARstart{T}{he} ever-ending pursuit of higher data rate sets the goal for the next-decade  wireless services: providing at least a challenging Tbit/s-level bit rate. A key advancement in 6G is changing the way of indoor communications from radio-frequency (RF) communications to free-space optical communications (FSO)~\cite{a180820.09}. Since the light frequency lies in the range of several hundred THz, which means very large bandwidth can be obtained, FSO has the potential to support Tbit/s-level data transmission.

\begin{figure}
	\centering
	\includegraphics[width=3.3in]{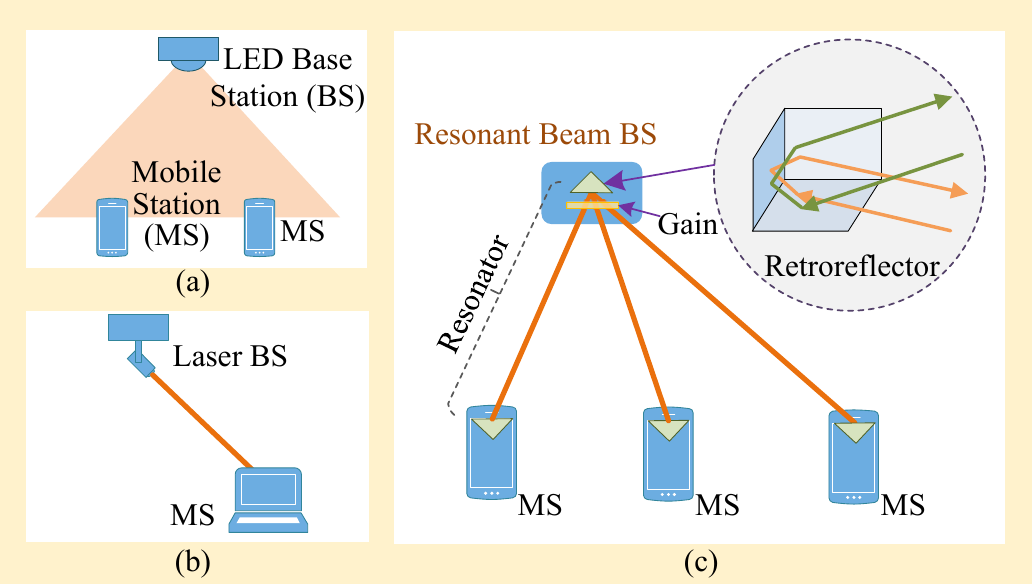}
	\caption{Comparison of communication technologies with: (a) non-directional radiation; (b) directional radiation; (c) retro-reflective resonant beam}
	\label{fig:principle}
\end{figure}

Existing FSO technologies include non-directional light (e.g., LED radiation) and directional light (e.g., laser beam).
As shown in Fig. \ref{fig:principle}(a), the light from the LED base station~(BS) is emitted at a large angle. Mobile stations~(MSs) within the coverage area can receive the light signal from the BS. However, the received light intensity decreases quickly as the distance grows. In a nutshell, non-directional light supports high mobility and multiple accessing but undergoes remarkable path loss~\cite{jovicic2013visible}. In contrast, directional light has higher power density and smaller path loss than non-directional light, as shown in Fig.~\ref{fig:principle}(b).
However, directional light needs positioning and beam steering to implement alignment and tracking of MSs, which highly demands response speed and accuracy~\cite{a190505.06}. Generally, fiber array can be used as the light source to generate directional beams; this design has achieved data rate of $10$~Gbit/s~\cite{a201130.02}. Through tilted fiber grating, light beam carrying $12$~Gbit/s date can be directed to the receiver at different location~\cite{a201130.03}. Spatial light modulator~(SLM) can be employed to generate hologram which is able to be controlled to cover specific receivers~\cite{a190514.01}. Rhee \textit{et al}. employed optical phase array~(OPA) to generate beams to arbitrary direction, and achieved the  data rate of $32$~Gbit/s~\cite{a201201.02}. Chun \textit{et al.} used micro-electro-mechanical system (MEMS) mirror to steer the light beam, and achieved the date rate of $35$~Gbit/s~\cite{a201201.03}. However, these technologies require positioning technology to acquire the accurate location of the receiver, which increases the costs and complexity. Besides, the mobility of these technologies is limited by the transmission delay of positioning signal, the refresh frequency of OPA, or the response speed of MEMS.
\begin{figure}
	\centering
	\includegraphics[width=3.1in]{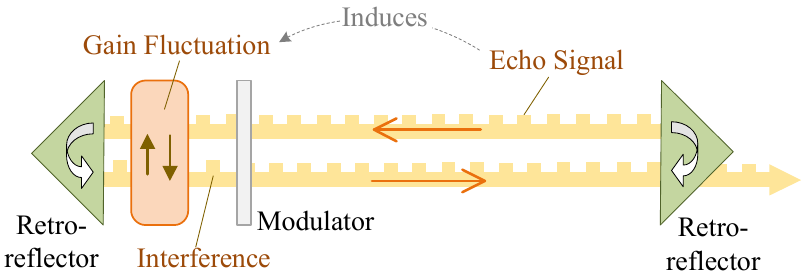}
	\caption{Echo interference issue in resonant beam communications system}
	\label{fig:problem}
\end{figure}
\begin{figure*}[t]
	\centering
	\includegraphics[width=5.5in]{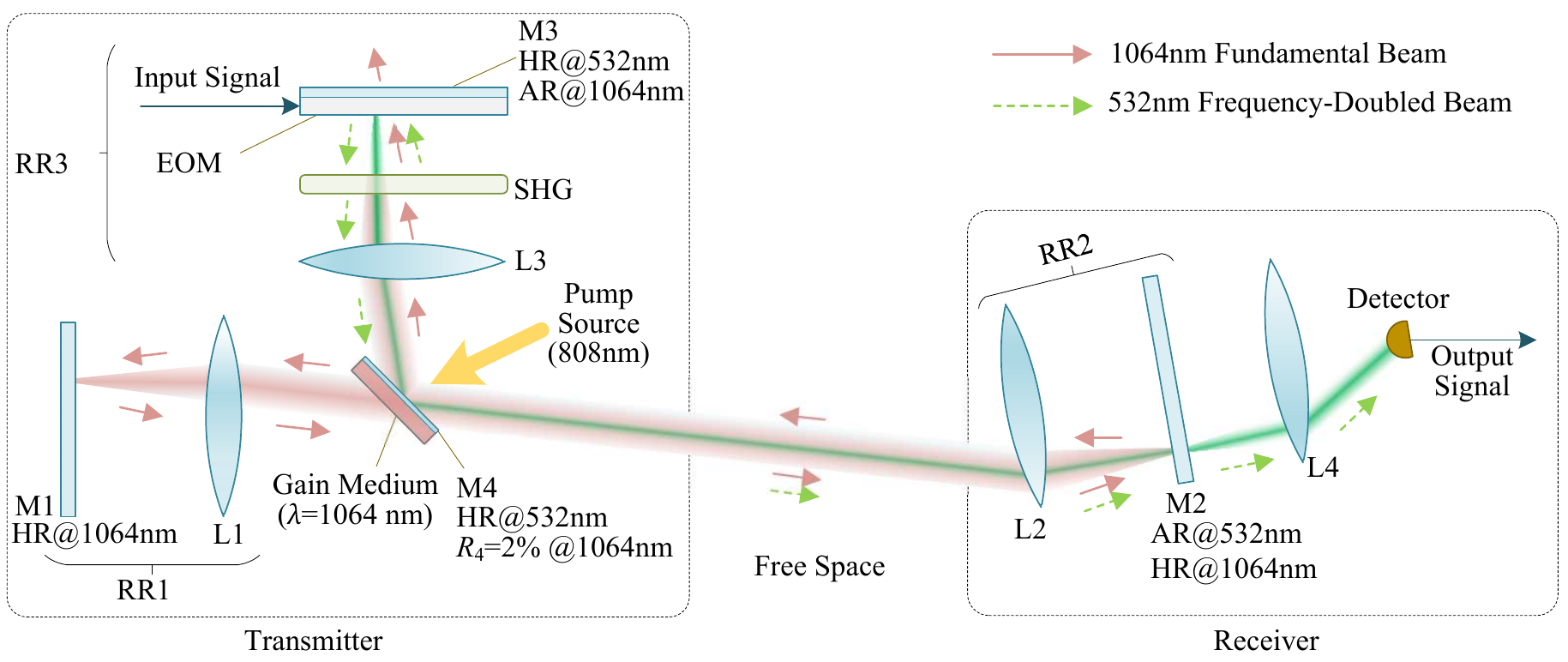}
	\caption{Echo-interference-free  design for resonant beam communications system with second harmonic generation (AR: anti-reflective coating, i.e., 100\% transmissivity; HR: high-reflective coating, i.e., 100\% reflectivity; L1, L2, L3, and L4 are lenses; M1, M2, M3, and M4 are flat mirrors; RR1, RR2, and RR3 are retroreflectors; EOM: electro-optical modulator; SHG: second harmonic generation medium; $R_4$: reflectivity of M4 at $1064$ nm)}
	\label{fig:design}
\end{figure*}

Figure~\ref{fig:principle}(c) depicts the resonant beam communications (RBCom) system, in which both the BS and the MS contain a retroreflector to reflect the incident lights back along the incident direction. Similar to the intelligent reflecting surface (IRS) enhanced communication system which adjusts the channel state to improve the  performance, RBCom channel has the adaptive ability in beamforming and alignment~\cite{a200524.02}. Two retroreflectors form a spatially distributed laser resonator naturally without the need of strict alignment,  providing RBCom with self-alignment feature and mobility~\cite{a180727.01, a190318.02,a190926.02}. In this retro-reflective resonator, photons generated by the gain medium through spontaneous emission; then, they travels back and forth and are enhanced by superposition. In this case, an intra-cavity resonant beam is spontaneously formed, connecting the BS and the MS.  The source power drives the gain medium to provide a power amplification capability. As the transmission loss is compensated by the gain, the beam power can reach a considerable level to support high SNR~\cite{a190213.02,a200528.01}.

However, the RBCom system undergoes echo interference. As in Fig.~\ref{fig:problem}, the transmitted signal is reflected back to the transmitter's gain medium and modulator by the receiver's retroreflector, affecting the gain stability and the subsequent modulation. This characteristic plus its solution relying on frequency-band shifting and optical filtering are presented in~\cite{a200508.02,a200528.02}. Nevertheless, even narrow-band optical filters are used, the modulator still has to operate at tens of GHz to ensure that the signal frequency lies out of the filter's passband. As high-speed modulation is supported by splitting the modulator's surface into small pixels, the driving power as well as the temperature will increase to a prohibitive level~\cite{a190926.04}. The demands for a practical and economic echo-interference-free RBCom systems motivate this work.

In this paper, to avoid echo interference, we exploit the fact that only the light wave in a certain frequency band affects the stimulation emission of the gain medium, and thus design a coupled optical path with second harmonic generation (SHG) to produce frequency-doubled carrier beam which lies out of the gain medium's stimulation frequency band. At the transmitter, the fundamental beam for SHG is extracted from the resonant beam formed within the spatially separated resonator through a coated mirror M4. At the receiver, the frequency-doubled beam passes through a partially reflective mirror M2 and finally arrives at the detector, while the low-frequency resonant beam is reflected to maintain the resonance. The coupled design allows the frequency-doubled beam transmitting along the free-space path of the resonant beam which automatically keeps pointing to the receiver at any time. The modulator only changes the frequency-doubled beam, so the resonant beam remains stable, avoiding echo interference.

The contributions of this work are as follows.
\begin{enumerate}
	\item[\bf 1)] We propose an SHG-based RBCom system design, which solves the echo interference problem of the original RBCom system, providing non-mechanical mobility and high capacity.
	\item[\bf 2)] We establish an analytical model for the SHG-based RBCom system, which can evaluate the beam radius, diffraction loss, beam power, and channel capacity, without the need of a numerical computing program that is usually adopted for evaluating such multi-lens resonator.
\end{enumerate}

The remainder of this paper is organized as follows.
Section~\ref{sec:model} details the model of the self-interference-free RBCom system. Section~\ref{sec:evalu} presents the performance evaluation of the proposed model. Section~\ref{sec:disc} discusses several issues for further research. Finally, conclusions
are drawn in Section~\ref{sec:con}.

\section{System Model}
\label{sec:model}

\subsection{System Design}
The SHG-based RBCom system consists of two coupled parts, i.e., the spatially separated laser resonator and the SHG path. A portion of the fundamental beam generated by the resonator is coupled into the SHG path and is converted into a frequency-doubled beam which serves as the communication carrier. After modulation, the frequency-doubled beam is coupled into the original fundamental beam path, and propagates to the receiver along this path.

As depicted in Fig.~\ref{fig:design}, RR1 and RR2 are retroreflectors with cat's eye structure, and are installed in the transmitter and the receiver, respectively. The spatially separated laser resonator is constituted by RR1, the gain medium, and RR2. The gain medium absorbs pump source light to enable optical amplification. Here Nd:YVO$_4$ crystal is employed as the gain medium, hence, the pump source is $808$-nm light~(usually generated by laser diodes; in addition to $808$~nm, some other wavelengths such as $880$ nm can also be used for pumping the Nd:YVO$_4$ crystal), and $1064\mbox{-nm}$ light can be amplified. According to the mechanism of laser resonators, $1064$-nm resonant beam can be generated in this resonator.

The surface on one side of the gain medium has a thin coating M4 which can reflect all the $532$-nm beam and a portion (e.g., reflectivity $R_4=2\%$) of $1064$-nm beam. Hence, the $1064$-nm resonant beam in the resonator is split by the partially reflective surface M4; a portion of the resonant beam is extracted as the fundamental beam for SHG; and the rest still stays in the resonating path between RR1 and RR2 to maintain the resonance. The $1064$-nm fundamental beam enters the retroreflector RR3 and passes the SHG medium to generate $532$-nm frequency-doubled beam.

The $532$-nm beam is modulated by the electro-optical modulator (EOM) and then reflected by mirror M3. Since RR1 and RR3 are designed to be mirror symmetric about M4, and they have capability of retro-reflection, the modulated $532$-nm beam can travel along the original path of the incident $1064\mbox{-nm}$  beam. After being reflected by M4, the modulated $532$-nm beam is coupled into the free-space propagating path of the resonant beam and received by the receiver. Mirror M2 is coated to reflect only $1064$-nm beam. Thus, the $532\mbox{-nm}$ beam can pass through M2. Finally, the $532$-nm beam is focused on the photon detector~(PD) by lens L4.

In RBCom, the modulated signals can pass through the mirror M2, without interference on the gain and the modulation operation. The fundamental beam is retro-reflected by M2 to maintain the resonance, which provides an automatically tracking path towards the receiver.

\subsection{Telecentric Cat's Eye Retroreflector}

\begin{figure}[t]
	\centering
	\includegraphics[width=2.8in]{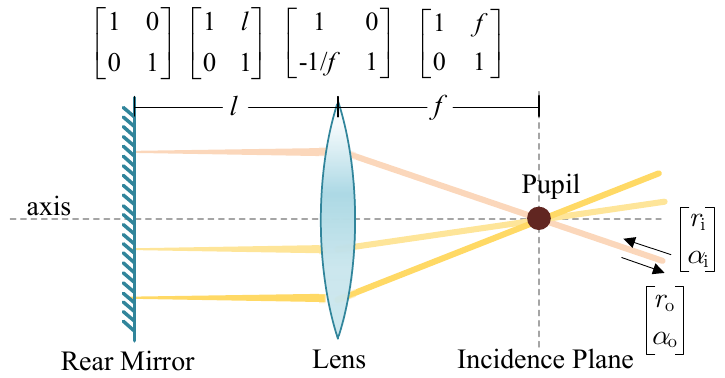}
	\caption{Telecentric cat's eye retroreflector}
	\label{fig:retroref}
\end{figure}

There are several kinds of retroreflectors, such as corner cube, cat's eye, and telecentric cat's eye, to constitute spatially separated laser resonators and enable the self-alignment features. In this work, we exploit the telecentric cat's eye retroreflector, since it has the following unique optical characteristics. As depicted in Fig.~\ref{fig:retroref}, the telecentric cat's eye consists of a flat rear mirror whose radius of curvature~(ROC) is infinite and a lens whose focal length is $f$. The interval between the mirror and the lens is $l$. A pupil locates at the right focal point of the lens. The light beam entering the pupil from arbitrary directions will: 1) be perpendicular to the rear mirror after passing through the lens; and 2) exit through the pupil along the original path of the incident beam.

Ray-transfer matrices are adopted to describe an optical system under the paraxial approximation. All the optical elements and the free spaces in the system are represented by individual matrices. The system ray-transfer matrix is the production of these individual matrices with the opposite order~\cite{a200522.04}. For the cat's eye shown in Fig.~\ref{fig:resonator}, the matrix is
\begin{align}
\mathbf{M}_{\rm RR}&=
\begin{bmatrix}  1&f\\ 0&1 \end{bmatrix}
\begin{bmatrix}  1&0\\ -1/f&1 \end{bmatrix}
\begin{bmatrix}  1&l\\ 0&1 \end{bmatrix}
\begin{bmatrix}  1&0\\ 0&1 \end{bmatrix} 	\nonumber\\
&~~~~
\begin{bmatrix}  1&l\\ 0&1 \end{bmatrix}
\begin{bmatrix}  1&0\\ -1/f&1 \end{bmatrix}
\begin{bmatrix}  1&f\\ 0&1 \end{bmatrix}\nonumber\\
&=
\begin{bmatrix}  1&0\\ -1/f_{\rm RR}&1 \end{bmatrix}
\begin{bmatrix}  -1&0\\ 0&-1 \end{bmatrix},
\label{equ:MRR}
\end{align}
where
\begin{equation}
f_{\rm RR}=1\bigg/\left(\dfrac{2l}{f^2}-\dfrac{2}{f}\right).
\end{equation}
From~\eqref{equ:MRR}, the matrix $\mathbf{M}_{\rm RR}$ is equivalent to the combination of an imaging device (negative identity  matrix) and a lens.
If $l=f$, the matrix $\mathbf{M}_{\rm RR}$ exhibits the characteristic of an ideal telecentric cat's eye; that is
\begin{equation}
\mathbf{M}_{\rm RR}\bigg|_{l=f}=
\begin{bmatrix}  -1&0\\ 0&-1 \end{bmatrix}.
\label{equ:idealRR}
\end{equation}
With the imaging characteristic of the ideal telecentric cat's eye depicted in~\eqref{equ:idealRR}, the output ray vector is expressed as
\begin{equation}
\begin{bmatrix}  r_{\rm o}\\ \alpha_{\rm o} \end{bmatrix}=
\mathbf{M}_{\rm RR}\bigg|_{l=f}
\begin{bmatrix}  r_{\rm i}\\ \alpha_{\rm i} \end{bmatrix}
=\begin{bmatrix}  -r_{\rm i}\\ -\alpha_{\rm i} \end{bmatrix},
\label{equ:inout-retro}
\end{equation}
where $r_{\rm o}$ ($r_{\rm i}$) is the transverse displacement of the ray from the optical axis at the incidence plane; and $\alpha_{\rm o}$ ($\alpha_{\rm i}$) is the slope of the output (input) ray. According to \eqref{equ:inout-retro}, the output ray is always parallel to the input ray. This imaging characteristic supports the mobility of the spatially distributed resonator, i.e., two retroreflectors don't need to be aligned strictly to each other, since the intra-cavity rays from one retroreflector can always be reflected back by the other retroreflector.

\subsection{Spatially Separated Resonator}

\begin{figure}[t]
	\centering
	\includegraphics[width=3.4in]{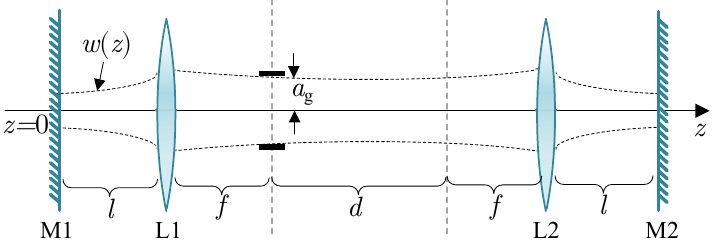}
	\caption{Schematic of intra-cavity beam radius distribution}
	\label{fig:resonator}
\end{figure}

The spatially separated resonator consists of two telecentric cat's eye retroreflectors, as in Fig.~\ref{fig:resonator}. ABCD matrices are used to analyze the stability of resonators and obtain the intensity distribution of transverse modes of the intra-cavity beam. The single-pass ABCD matrix of the resonator in Fig.~\ref{fig:resonator} which describes the ray traveling from mirror M1 to mirror M2 is obtained as~\cite{a200224.01,a190511.01,a181221.01}
\begin{align}
\begin{bmatrix}  A&B\\ C&D \end{bmatrix} &=
\begin{bmatrix}  1&0\\ 0&1 \end{bmatrix}
\begin{bmatrix}  1&l\\ 0&1 \end{bmatrix}
\begin{bmatrix}  1&0\\ -\frac{1}{f}&1 \end{bmatrix}
\begin{bmatrix}  1&2f+d\\ 0&1\end{bmatrix} \nonumber\\
&~~~~
\begin{bmatrix}  1&0\\ -\frac{1}{f}&1 \end{bmatrix}
\begin{bmatrix}  1&l\\ 0&1 \end{bmatrix}
\begin{bmatrix}  1&0\\ 0&1 \end{bmatrix}\nonumber\\
&=\begin{bmatrix}  -1-\dfrac{d}{f}+\dfrac{dl}{f^2}&2f-2l+d-\dfrac{2dl}{f}+\dfrac{dl^2}{f^2}\\\dfrac{d}{f^2}&-1-\dfrac{d}{f}+\dfrac{dl}{f^2} \end{bmatrix}.
\label{equ:ABCD}
\end{align}

One can calculate the beam radius at the mirrors located at both ends of the resonator by the theory of equivalent resonator.
From~\eqref{equ:ABCD}, the equivalent g-parameters \{$g_{1}^*$, $g_{2}^*$\} and the equivalent length $L^*$ of the resonator is~\cite{a181221.01}
\begin{align}
g_1^*&=A=\frac{d}{2f_{\rm RR}}-1, \label{equ:g1g1}\\
g_2^*&=D=\frac{d}{2f_{\rm RR}}-1,\\	
L^*&=B=-\frac{f^2}{f_{\rm RR}}+\frac{df^2}{4f_{\rm RR}^2}.
\label{equ:g1g1L}
\end{align}
From~\eqref{equ:g1g1}-\eqref{equ:g1g1L}, the resonator can be described by three parameters \{$d$, $f$, $f_{\rm RR}$\}. The value of $l$ can be obtained if $f$ and $f_{\rm RR}$ are known.
The resonator is stable if the  condition $0<g_1^*g_2^*<1$ holds~\cite{a181221.01}. Therefore, we can derive the following stability condition for our resonator, namely
\begin{equation}
0\le d<4f_{\rm RR}~\mbox{and}~d\neq 2f_{\rm RR}.
\end{equation}
Note that $d=2f_{\rm RR}$ is also acceptable, where the equivalent resonator is a confocal cavity which exhibits the best stability. We found that the condition $l>f$ (i.e., $f_{\rm RR}>0$) should be satisfied when two retroreflectors have the same configuration, so that $\mathbf{M}_{\rm RR}$ exhibits the capability of focusing light and the resonator can be stable. 

According to laser principle, the fundamental mode, TEM$_{00}$, in a stable resonator is a Gaussian beam. The TEM$_{00}$ mode radius at mirror M1 is obtained as~\cite{a181221.01,a200515.04}
\begin{equation}
w_{00}(0)=\sqrt{\frac{\lambda \left|L^*\right|}{\pi}\sqrt{\frac{g_2^*}{g_1^*(1-g_1^*g_2^*)}}},
\end{equation}
where $\lambda$ is the wavelength of the beam, and $\pi$ is the ratio of a circle's circumference to its diameter.

The parameters of the Gaussian beam propagating inside the resonator can be obtained according to the fact that mirror surfaces  are the constant-phase surfaces for the beam~\cite{a181221.01}. Since the ROC of M1 is infinite, the ROC of the constant-phase surface of the beam at M1 is also infinite, which means the waist of the beam between M1 and L1 locates at M1, as only the constant-phase surface at the waist of the Gaussian beam exhibits infinite ROC. Once we obtain the waist radius of the Gaussian beam, we can obtain its g-parameter as a function of the location along the $z$-axis. Then, we can compute the mode radius at any location inside or outside the resonator. Finally, the $q$-parameter at M1 is~\cite{a181221.01}
\begin{equation}
q_0=j\dfrac{\pi w_{00}^2(0)}{\lambda},
\end{equation}
where $j:=\sqrt{-1}$. By using the ray-transfer matrix, i.e. the ABCD law~\cite{a181221.01}, the $q$-parameter along the axis is obtained as
\begin{equation}
q(z)=
\left\{
\begin{aligned}
&q_0+z, ~~~~~~~~~~~~~~~~~~~~~~~~~~z\in[0,z_{\rm L1}]\\
&\frac{q(z_{\rm L1})}{{-q(z_{\rm L1})}/{f}+1}+(z-z_{\rm L1}), z\in(z_{\rm L1},z_{\rm L2}]\\
&\frac{q(z_{\rm L2})}{{-q(z_{\rm L2})}/{f}+1}+(z-z_{\rm L2}), z\in(z_{\rm L2},z_{\rm M2}]\\
\end{aligned}
\right.
\label{equ:qParam}
\end{equation}
where $z_{\rm L1}=l$, $z_{\rm L2}=l+2f+d$, and $z_{\rm M2}=2l+2f+d$. The Gaussian beam radius $w_{00}(z)$ at the location $z$ is determined  by the imaginary parts of $1/q(z)$, namely~\cite{a181221.01}
\begin{equation}
w_{00}(z)=\sqrt{-\frac{\lambda}{\pi \Im[1/q(z)]}},
\end{equation}
where $\Im[\cdot]$ takes the imaginary part of a complex quantity. 


Next, we consider the higher-order transverse modes, TEM$_{mn}$, in the resonator. These higher-order modes have a larger radius than TEM$_{00}$ mode. All these modes are superposed together, forming an intra-cavity resonant beam. The beam radius $w(z)$ at an arbitrary
location is calculated by~\cite{a181221.01}
\begin{equation}
w(z)= w_{00}(z)M,
\label{equ:wz}
\end{equation}
where $M$ is the beam propagation factor. Since the gain medium aperture is much smaller than that of other mirrors and lens in the cavity, the beam propagation factor can, to a good approximation, be calculated with~\cite{a181221.01}
\begin{equation}
M=\dfrac{a_{\rm g}}{w_{00}(l+f)},
\label{equ:Msqr}
\end{equation}
where $a_{\rm g}$ is the radius of the gain medium aperture, and $w_{00}(l+f)$ is the radius of TEM$_{00}$ mode at the location of the gain medium. The mode index $m$ and $n$ are limited by~\cite{a181221.01}
\begin{equation}
2m+n+1 \leqslant  M^2.
\label{equ:mn}
\end{equation}

\subsection{Fundamental Beam Generation}

\begin{figure}
	\centering
	\includegraphics[width=3.3in]{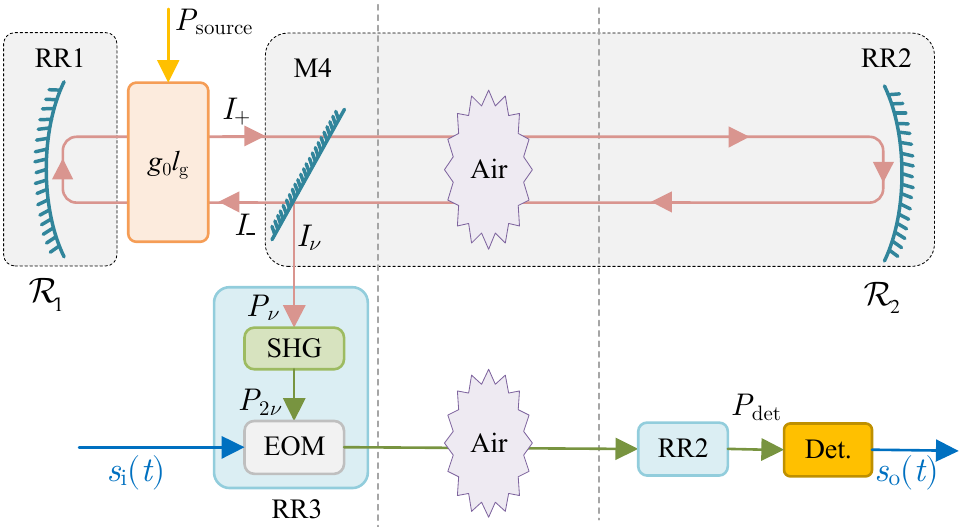}
	\caption{Power and signal flow (Det. is the detector)}
	\label{fig:model}
\end{figure}

As depicted in Fig.~\ref{fig:model}, photons oscillate in the  resonator and are amplified by the gain medium, forming the resonant beam. A portion of the resonant beam is extracted by M4 as the fundamental beam for SHG. In this part, we create the mathematical model to calculate the power of the extracted fundamental beam.

The resonance can be built when the threshold condition is met; that is~\cite{a181218.01}
\begin{equation}
2g_0l_{\rm g}\geqslant\delta,
\end{equation}
where $g_0$ is the small-signal gain coefficient of the gain medium, $l_{\rm g}$ is the path length of the beam internal the gain medium, and $\delta$ is the total round-trip loss coefficient of the resonator. The small-signal gain coefficient is determined by the power density of the pump light absorbed by the gain medium, namely~\cite{a181218.01}
\begin{equation}
g_0=\dfrac{\eta_c P_{\rm in}}{I_{\rm s} V},
\label{equ:g0}
\end{equation}
where $I_{\rm s}=(h\nu)/(\sigma_s\tau_f)$ is the saturation intensity; $\nu=c/\lambda$ is the wavelength of the intra-cavity beam; $\sigma_{\rm s}$ and $\tau_{\rm f}$ are the stimulated emission cross section and the fluorescence time of the gain medium, respectively; $P_{\rm in}$ is the electrical power used for driving the pump laser diodes; and $V$ is the volume of the gain medium. The combined pumping efficiency $\eta_c$ is expressed as~\cite{a181218.01}
\begin{equation}
\eta_c=\eta_{\rm P}\eta_{\rm t} \eta_{\rm a} \eta_{\rm Q} \eta_{\rm S} \eta_{\rm B},
\end{equation}
where $\eta_{\rm P}$ is the pump source generation efficiency, $\eta_{\rm t}$ is the pump source transfer efficiency, $\eta_{\rm a}$ is the pump source absorb efficiency, $\eta_{\rm Q}$ is the quantum efficiency, $\eta_{\rm S}=\nu/\nu_{\rm p}$ is the Stokes factor, $\nu_{\rm p}$ is the pump light frequency, and $\eta_{\rm B}$ is the overlap efficiency defined as the ratio of the mode volume of the resonant beam to the volume of the active gain medium.

The losses in the resonator include reflection loss, transmission loss, absorption loss, scattering loss, and diffraction loss. These losses are induced by  mirrors, lenses, the gain medium, and the air-transmission. The total round-trip loss coefficient is defined as
\begin{equation}
\delta=-\ln\left(\Gamma_{\rm RR1} \Gamma_{\rm diff} \Gamma_{\rm g} \Gamma_{\rm M4} \Gamma_{\rm air}\Gamma_{\rm RR2}\right),
\end{equation}
where $\Gamma_{\rm RR1}$, $\Gamma_{\rm RR2}$,  $\Gamma_{\rm g}$, and $\Gamma_{\rm air}$ are the round-trip loss factors induced by  RR1, RR2, the gain medium, and the air, respectively; $\Gamma_{\rm M4}=1-R_4$ is the transmissivity of M4 at $1064$~nm; $\Gamma_{\rm diff}$ is the round-trip diffraction loss factor. The round-trip air-transmission loss factor is expressed as~\cite{a200427.04}
\begin{equation}
\Gamma_{\rm air}(d)=\mathrm{e}^{-2\alpha_{\rm air} d},
\end{equation} where $\alpha_{\rm air}=10^{-4}$~m$^{-1}$ is the loss coefficient of  clear air, and $d$ is the transmission distance.

Next, we compute the power of the fundamental beam extracted by M4. According to Rigrod analysis, for the simplest resonator with only two mirrors and one gain medium lied between the two mirrors, the intensity of the rightward-traveling light at the right mirror is computed by~\cite{a181224.01}
\begin{equation}
I_{+}=\frac{I_{\rm s}}{(1+r_2/r_1)(1- r_1 r_2)}\left[ g_0 l_{\rm g} - \ln\frac{1}{r_1 r_2}\right],
\label{equ:I+}
\end{equation}
where $r_1\equiv \sqrt{\mathcal{R}_1}$ and $r_2 \equiv \sqrt{\mathcal{R}_2}$ are the voltage reflection coefficients of the mirrors; and $\mathcal{R}_1$ and $\mathcal{R}_2$ are the reflectivity of the mirrors.

In our system model, we combine all the loss factors into two equivalent reflectance $\mathcal{R}_1$ and $\mathcal{R}_2$, as demonstrated in Fig.~\ref{fig:model}.
Considering the receiver's orientation angle \{$\alpha$, $\beta$\} and attitude angle \{$\theta$, $\phi$\}~\cite{a201125.02}, we use $\mathcal{R}_1(\bm\zeta,d)$ to denote the equivalent reflectivity for the combined loss induced by RR1, the diffraction loss, and the gain medium; that is
\begin{equation}
\mathcal{R}_1(\bm{\zeta},d)=\Gamma_{\rm RR1}(\bm{\zeta})\Gamma_{\rm g}(\bm{\zeta})\Gamma_{\rm diff}(\bm{\zeta},d),
\end{equation}
where $\bm\zeta=[\alpha,\beta,\theta,\phi]$, and $d$ is the transmission distance. $\mathcal{R}_2(\bm\zeta)$ denotes the equivalent reflectivity for the combined loss induced by M4, the air, and RR2; that is
\begin{equation}
\mathcal{R}_2(\bm{\zeta},d)=\Gamma_{\rm M4}(\bm{\zeta})\Gamma_{\rm air}(d)\Gamma_{\rm RR2}(\bm{\zeta}).
\end{equation}
The intensity of the leftward-traveling light on the left surface of M4 is expressed as
\begin{equation}
I_{-}=\mathcal{R}_2(\bm{\zeta},d) I_{+}.
\end{equation}
Then the intensity of the fundamental beam  extracted by M4 is obtained as
\begin{equation}
I_{\nu}(\bm{\zeta},d)=\frac{R_4(\bm{\zeta})}{1-R_4(\bm{\zeta})}I_{-}=\frac{\mathcal{R}_2(\bm{\zeta},d) R_4(\bm{\zeta})}{1-R_4(\bm{\zeta})} I_{+},
\label{equ:Inu1}
\end{equation}
where $R_4(\bm\zeta)$ is the reflectivity of M4 at the frequency $\nu$.
From~\eqref{equ:g0},~\eqref{equ:I+}, and~\eqref{equ:Inu1}, the  average power of the extracted fundamental beam is then obtained as
\begin{align}
P_{\nu}(\bm{\zeta},d)&=A_{\rm g}(\bm{\zeta}) I_{\nu}(\bm{\zeta},d) \nonumber\\
&= \eta_\nu(\bm{\zeta},d)[P_{\rm in}- P_{\rm th}(\bm{\zeta},d)]
\label{equ:pnu}
\end{align} 
where
\begin{equation}
\eta_\nu(\bm{\zeta},d)=\frac{l_{\rm g}(\bm{\zeta}) \eta_c(\bm{\zeta},d)  A_{\rm g}(\bm{\zeta})\mathcal{R}_2(\bm{\zeta},d) R_4(\bm{\zeta})  }{ V\left[1-R_4(\bm{\zeta})\right]\left[1+\frac{r_2(\bm{\zeta},d)}{r_1(\bm{\zeta},d)}\right]\left[1- r_1(\bm{\zeta},d)r_2(\bm{\zeta},d)\right]},
\label{equ:eta-nu}
\end{equation}
and
\begin{equation}
P_{\rm th}(\bm{\zeta},d)=\ln\left[\frac{1}{r_1(\bm\zeta,d) r_2(\bm\zeta,d)}\right]\cdot\dfrac{I_{\rm s} V}{l_{\rm g}(\bm\zeta) \eta_c(\bm\zeta)  }.
\end{equation}
Here $A_{\rm g}(\bm\zeta)$ represents the beam cross section area at the gain medium, and $l_{\rm g}(\bm\zeta)$ is the path length of the beam internal the gain medium. Note that, among all the efficiency components of $\eta_{\rm c}(\bm\zeta,d)$, only the overlap efficiency $\eta_{\rm B}(\bm\zeta,d)$ is relevant to  $\bm\zeta$ and $d$.  We only consider multi-mode oscillation, as the apertures of the devices in the cavity are much greater than the cross section of the TM$_{00}$ mode; and on this condition, the beam can be assumed to have uniform transversal intensity pattern. Therefore, we neglect the parameter $\bm\zeta$ and $d$, and use $\eta_{\rm c}$ and $\eta_{\rm B}$ to denote the combined pumping efficiency and the overlap efficiency, respectively.

Well-designed AR/HR coating has extremely high transmittance/reflectivity within a large angle of incidence (generally, $> 99$\%, within $\pm10^\circ$). The diffraction loss depends on the location of the receiver. The diffraction loss can be calculated by numerical simulation program (for example, the Fox-Li algorithm)~\cite{a201125.01}. In the following, we only consider the special case, where $\bm\zeta= [0,0,0,0]$, because the optimum performance is achieved at the direction along the optical axis. According to~\eqref{equ:pnu}, the  closed-form  propagation model can be expressed as
\begin{align}
P_{\nu}(d)&= \eta_\nu(d)[P_{\rm in}- P_{\rm th}(d)]
\label{equ:qu:pnu-simp}
\end{align} 
For calculating the explicit fundamental beam power expressed by~\eqref{equ:qu:pnu-simp}, the diffraction loss can be computed by the model presented in Appendix~\ref{sec:diffLossAppr}. For simplicity, we also assume optimum shape of gain medium. In this case, the radius of beam cross section at the gain medium  $A_{\rm g}=\pi a_{\rm g}^2$, where $a_{\rm g}$ is the radius of the gain medium aperture. Besides, the beam path length internal the gain medium $l_{\rm g}$ is approximate to the gain medium thickness.

\subsection{Second Harmonic Generation}

The fundamental beam passes through lens L3 and then enters the SHG medium. After that, the frequency-doubled beam is generated. SHG is a nonlinear optical process, in which two photons with the same frequency interact with the SHG medium to generate a new photon with twice the frequency and energy of the original photon. Some birefringent crystals are capable of generating frequency-doubled beam, such as potassium titanyl phosphate (KTP), potassium dihydrogen phosphate (KDP), and  lithium niobate~(LiNbO$_3$).

The SHG efficiency is determined by the input fundamental beam intensity (power $P_\nu$ divided by beam cross section $A_{\nu}$) and the length of the SHG medium $l_s$. Here we assume the fundamental beam is a plane-wave beam with critical phase matching and the effect of walk-off is neglected. Thus, the power of the frequency-doubled beam is obtained as~\cite{a181218.01, a200503.01}
\begin{equation}
P_{2\nu}=  \dfrac{Kl_{\rm s}^2 P_\nu^2}{A_{\nu}},
\label{equ:P2nu}
\end{equation}
where
\begin{equation}
K=\dfrac{8 \pi^2 d_{\rm eff}^2}{\varepsilon_0 c \lambda^2 n_0^3},
\end{equation}
$d_{\rm eff}$ is the efficient nonlinear coefficient of the SHG medium, $\varepsilon_0$ is the vacuum permeability, $c$ is the light speed, $n_0$ is the
refractive index of the SHG medium, and $A_{\nu}$ is the cross section area of the fundamental beam that enters the SHG medium. Supposing the output surface of SHG medium is attached to M3, and neglecting the thickness of the EOM, we obtain $A_{\nu}=\pi w^2(l_{\rm s})$ according to~\eqref{equ:wz}. Note that the plane-wave approximation is valid only when the length of SHG medium $l_s$ is smaller than the Rayleigh length ($z_{\rm R}=\pi w_{00}^2(0)/\lambda$) of the fundamental beam.

For each incident frequency, a specific phase matching angle should be satisfied for SHG progress. Otherwise, the conversion efficiency would be extremely low. Therefore, in our system, the SHG effect is neglected for the modulated beam  reflected back by M3, as the phase matching angle is set specifically for the fundamental frequency.

\subsection{Communication Channel Model}

The frequency-doubled beam is modulated by the EOM to carry information. Then the modulated beam travels along the path of the original fundamental beam to be received by the PD. The received optical power at the PD is
\begin{equation}
\hat{P}_{\rm r}=\eta_{\rm dev}\Gamma_{\rm air}^{1/2}P_{2\nu},
\end{equation}
where $\eta_{\rm dev}$ is the combined transmission efficiency of all the devices that the modulated beam passes through, including EOM, the SHG medium, L1, M4, RR2, L4, and the PD. Note that $\hat{P}_{\rm r}$ represents the peak  power of the optical signal. 

Wireless optical communications can be modeled as a linear time-invariant system~\cite{a190505.06}. Let $s_{\rm i}(t)$ denote the source signal to be transmitted. The detected current signal at the receiver is expressed as
\begin{equation}
s_{\rm o}(t)= \gamma P_{\rm 2\nu}s_{\rm i}(t)*h(t)+n(t),
\label{equ:Sof}
\end{equation}
where $*$ is the convolution operator, $\gamma$ is the PD's responsivity, and $n(t)$ denotes the additive white Gaussian noise~(AWGN). $h(t)$ is the channel impulse response of the transmission from the modulator to the detector, which is expressed as
\begin{equation}
h(t)=h_{\rm EOM}(t)*h_{\rm air}(t)*h_{\rm det}(t),
\label{equ:ht}
\end{equation}
where $h_{\rm EOM}(t)$, $h_{\rm air}(t)$, and $h_{\rm det}(t)$ are the impulse response functions of the EOM, the air-transmission channel and the PD, respectively. The effects on frequency domain imposed by the optics and air transmission channel can be neglected, as the bandwidth of the baseband signal is very narrow compared with the light frequency.

The noises include the pump source noise, the background light noise, the PD noise, and the demodulating noise. However, since the gain medium acts like a capacitor, only low-frequency noise from the pump source can be transferred to the receiver, which is out of the communication band and can be easily filtered. The background light noise can be filtered by an optical filter. Besides, well-designed demodulation circuits can dramatically reduce the noise and amplify the signal. Only the Gaussian white noise generated by photo-detector is considered, because the receiving power is great so that the shot noise and the thermal noise of the photodetector are dominant among these noise factors.

According to the classic literature, the channel capacity of our system is computed by~\cite{a200416.03}
\begin{equation}
\tilde{C}=\log_2\left(1+\textit{SNR}\right),
\end{equation}
where the signal-to-noise ratio $\textit{SNR}$  is obtained as
\begin{equation}
\textit{SNR}=\dfrac{(\gamma P_{\rm r})^2}{\sigma_{\rm total}^2},
\end{equation}
where $P_{\rm r} < \hat{P}_{\rm r}$ is the average received optical signal power. The significant noises power includes the thermal noise power and the shot noise power of the PD; that is~\cite{a200424.01}
\begin{equation}
\sigma_{\rm total}^2=\sigma_{\rm shot}^2+\sigma_{\rm thermal}^2.
\end{equation}
The shot noise power (A$^2$) is given by
\begin{equation}
\sigma_{\rm shot}^2=2e(\gamma P_{\rm r} + I_{\rm bk}) B,
\end{equation}
where $e$ is the electron charge, $I_{\rm bk}=5100~\mu$A is the background radiation induced photocurrent~\cite{a200427.02}, and $B$ is the bandwidth of PD.
The thermal noise power (A$^2$) is given by
\begin{equation}
\sigma_{\rm thermal}^2=\frac{4kT B}{R_{\rm L}},
\end{equation}
where  $k$ is the Boltzmann constant, $T$ is the absolute temperature, and $R_{\rm L}$ is the load resistor.

\begin{figure}[t]
	\centering
	\includegraphics[width=3.4in]{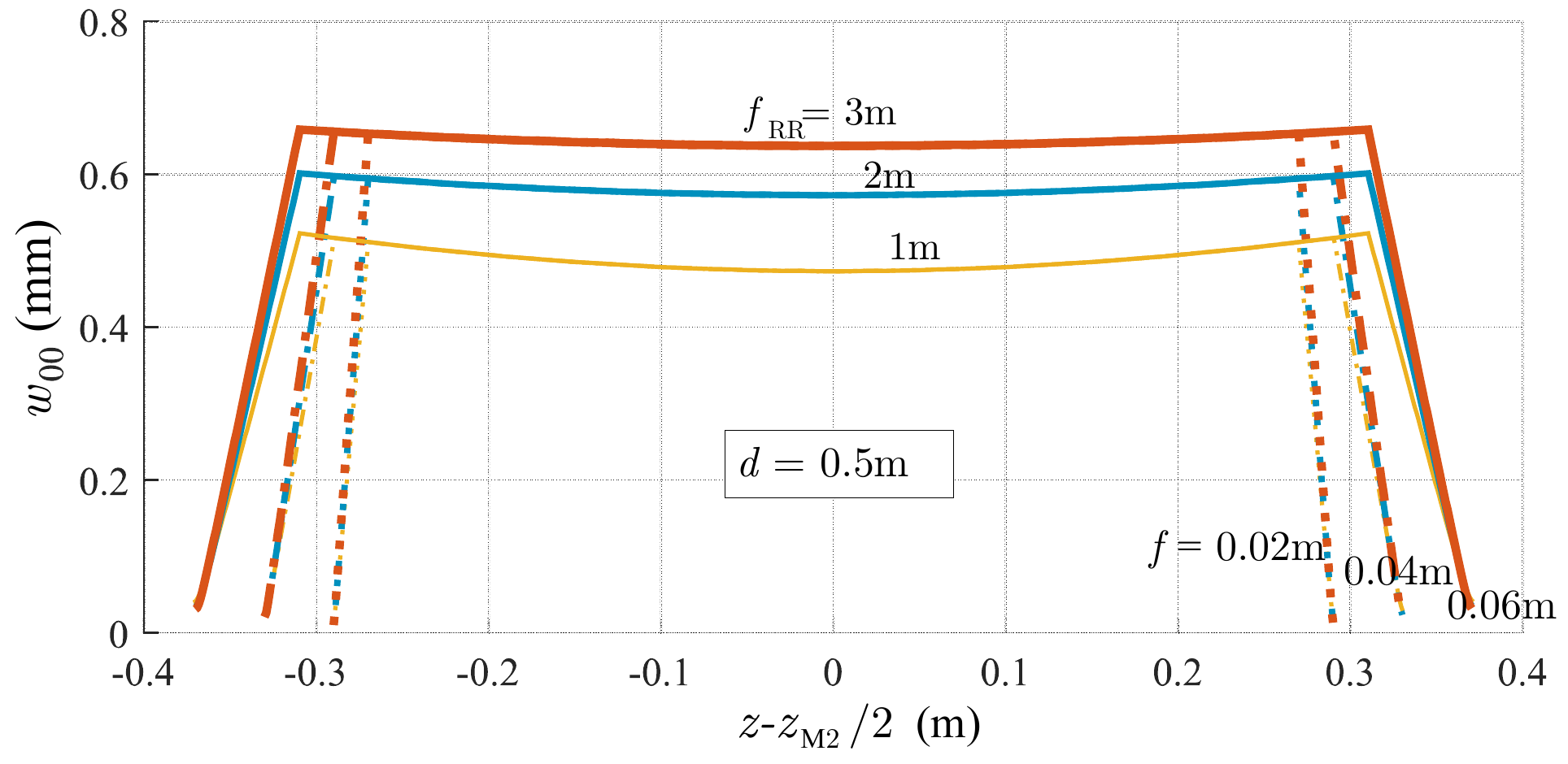}
	\caption{Beam radius of TEM$_{00}$ mode along the resonator's $z$ axis for different $f_{\rm RR}$ and focal length $f$ ($z_{\rm M2}=2l+2f+d$ is the resonator's length for each case)}
	\label{fig:br-fRR}
\end{figure}

\begin{figure}[t]
	\centering
	\includegraphics[width=3.4in]{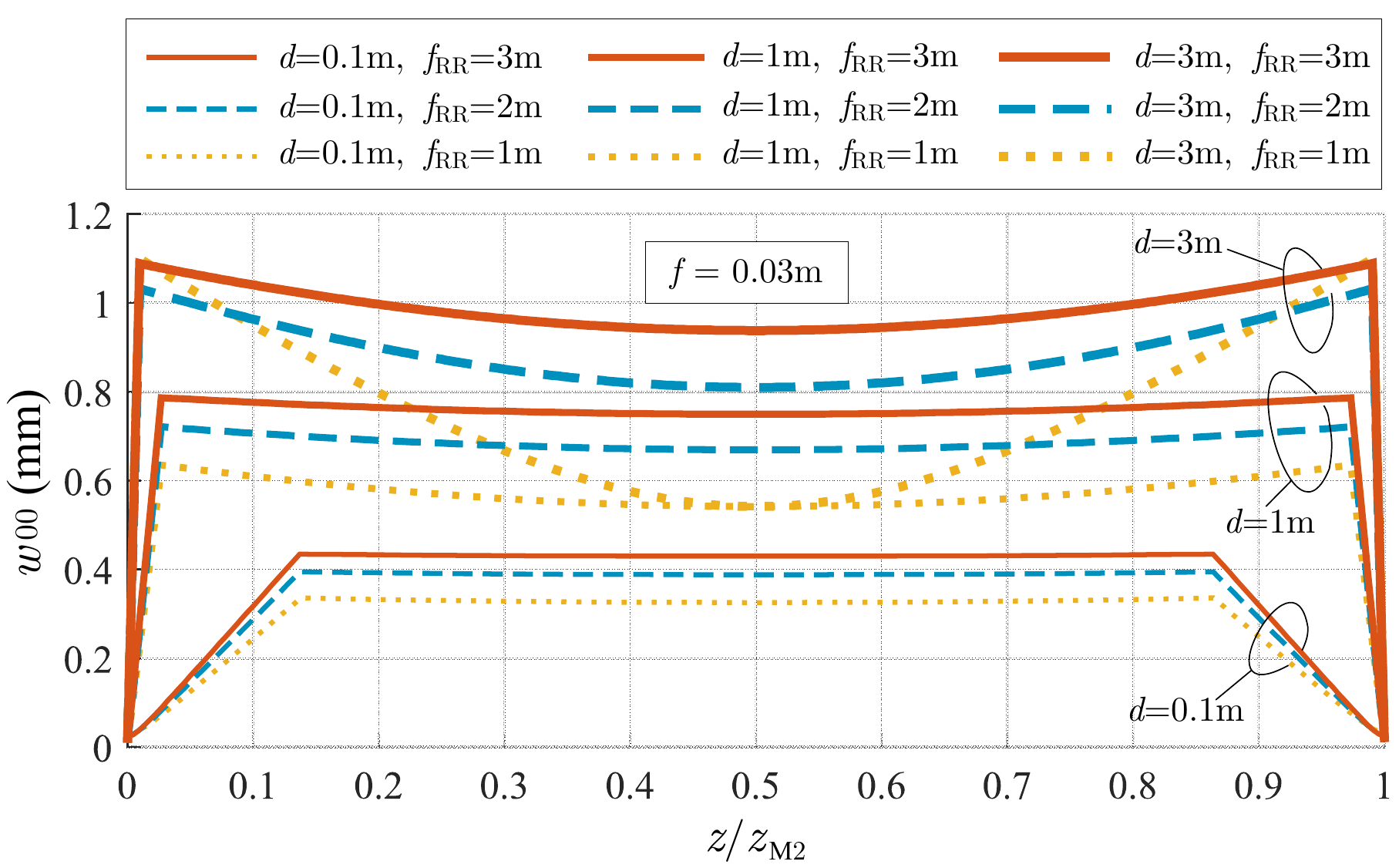}
	\caption{Beam radius of TEM$_{00}$ mode along the resonator's $z$ axis for different $f_{\rm RR}$ and distance $d$ ($z/z_{\rm M2}$ is the normalized $z$-coordinate, where $z_{\rm M2}=2l+2f+d$ is the resonator's length for each case)}
	\label{fig:br-d}
\end{figure}

\begin{table} [t] 
	\caption{Resonator Parameters~\cite{a181218.01,a200527.01}}
	\renewcommand{\arraystretch}{1.2}
	\centering
	\begin{tabular}{ l l l}
		\hline
		\textbf{Parameter} & \textbf{Symbol} &  \textbf{Value} \\
		\hline
		Stimulated emission cross section & $\sigma$ & $15.6\times10^{-23}$ m$^{2}$\\
		Fluorescence lifetime&$\tau_f$ & 100 $\upmu$s\\
		Fundamental beam wavelength & $\lambda$ & $1064$ nm\\
		Radius of gain medium aperture  & $a_{\rm g}$ & $3$ mm \\
		Gain medium thickness & $l_{\rm g}$ & $1$ mm\\
		Quantum efficiency& $\eta_{\rm Q}$ & $95\%$\\
		Stocks factor& $\eta_{\rm S}$ & $76\%$\\
		Overlap efficiency& $\eta_{\rm B}$ & $90\%$\\
		Pump source efficiency & $\eta_{\rm P}$ &75\%\\
		Pump source transfer efficiency&$\eta_{\rm t}$& 99\%\\
		Absorb efficiency&$\eta_{\rm a}$&$91\%$\\
		\hline
	\end{tabular}
	\label{tab:paramReson}
\end{table}
\begin{table} [t]
	\caption{Second Harmonic Generation Parameters~\cite{a181218.01}}
	\renewcommand{\arraystretch}{1.2}
	\centering
	\begin{tabular}{ l l l}
		\hline
		\textbf{Parameter} & \textbf{Symbol} &  \textbf{Value} \\
		\hline
		Efficient nonlinear coefficient & $d_{\rm eff}$ & $4.7$~pm/V\\
		refractive index&$n_0$ & 2.23 \\
		SHG medium length & $l_{\rm s}$ & $2$ mm\\
		\hline
	\end{tabular}
	\label{tab:paramSHG}
\end{table}

\section{Performance Evaluation}\label{sec:evalu}
\label{sec:perf}

In this section, we analyze the beam radius, the power of the fundamental beam, and the channel capacity of the proposed RBCom system. We set $P_{\rm r}=\hat{P}_{\rm r}$ to obtain the upper bound of the achievable channel capacity. The gain medium is an Nd:YVO$_4$ crystal. The parameters of the resonator are listed in Table~\ref{tab:paramReson}. Considering $\bm\zeta= [0,0,0,0]$, the loss factors $\Gamma_{\rm RR1}$, $\Gamma_{\rm RR2}$, $\Gamma_{\rm g}$, and $\eta_{\rm dev}$ are set to $1$. An LiNbO$_3$ crystal is employed as the SHG medium, as it has high nonlinear coefficient and $0^\circ$ walk-off angle at $1064$~nm; its parameters are specified in  Table~\ref{tab:paramSHG}. Besides, we consider the channel bandwidth is determined by the bandwidth of the modulator, thus, we set $B=800$~MHz~\cite{a200520.01}. The PD's responsivity $\gamma=0.6$~A/W. The temperature  $T=295$~K. The load resistor connected to the PD is $R_{\rm L}=10$~k$\Omega$~\cite{a200424.01,a200426.01}.

\subsection{Beam Radius}
Figure \ref{fig:br-fRR} shows  the beam radius $w_{00}$ of the TEM$_{00}$ mode distributed along the resonator's axis for the transmission distance $d = 0.5$~m. The beam radius are symmetric about the half-distance position $z=z_{\rm M2}/2$. Besides, $w_{00}$ reaches the minimum on mirrors M1 and M2, and reaches the maximum on lenses L1 and L2. The beam waist (the position where the Gaussian beam reaches the smallest radius) in the free space is located at the half-distance position. Moreover, as $f_{\rm RR}$ increases, the beam radius in the free space increases. Besides, the focal length $f$ has less effect on the beam radius profile in the free space if $f_{\rm RR}$ is specified.

We also analyzed the relationship between the beam radius profile and the resonator parameters including $d$ and $f_{\rm RR}$, for a certain $f=0.03$~m. In Fig. \ref{fig:br-d}, with the same $d$, a greater $f_{\rm RR}$ results in a larger beam waist $w_{00}$ in the free space.
With a given $f_{\rm RR}$, as distance $d$ increases, the beam radius $w_{00}$ near the lens and the gain medium becomes larger. This analysis reveals the fact that the intra-cavity beam is always focused onto the receiver under the stable state, so the received power as well as the channel capacity can reach a considerable level. As depicted in~\eqref{equ:mn}, a larger TEM$_{00}$ mode radius $w_{00}$ leads to a smaller mode number, which indicates a higher diffraction loss, a more inhomogeneous transverse intensity distribution, and a smaller overlap efficiency. Therefore, a gain medium with large cross section is required if we design a long-distance communication system.

\begin{figure} [t]
	\centering
	\includegraphics[width=3.4in]{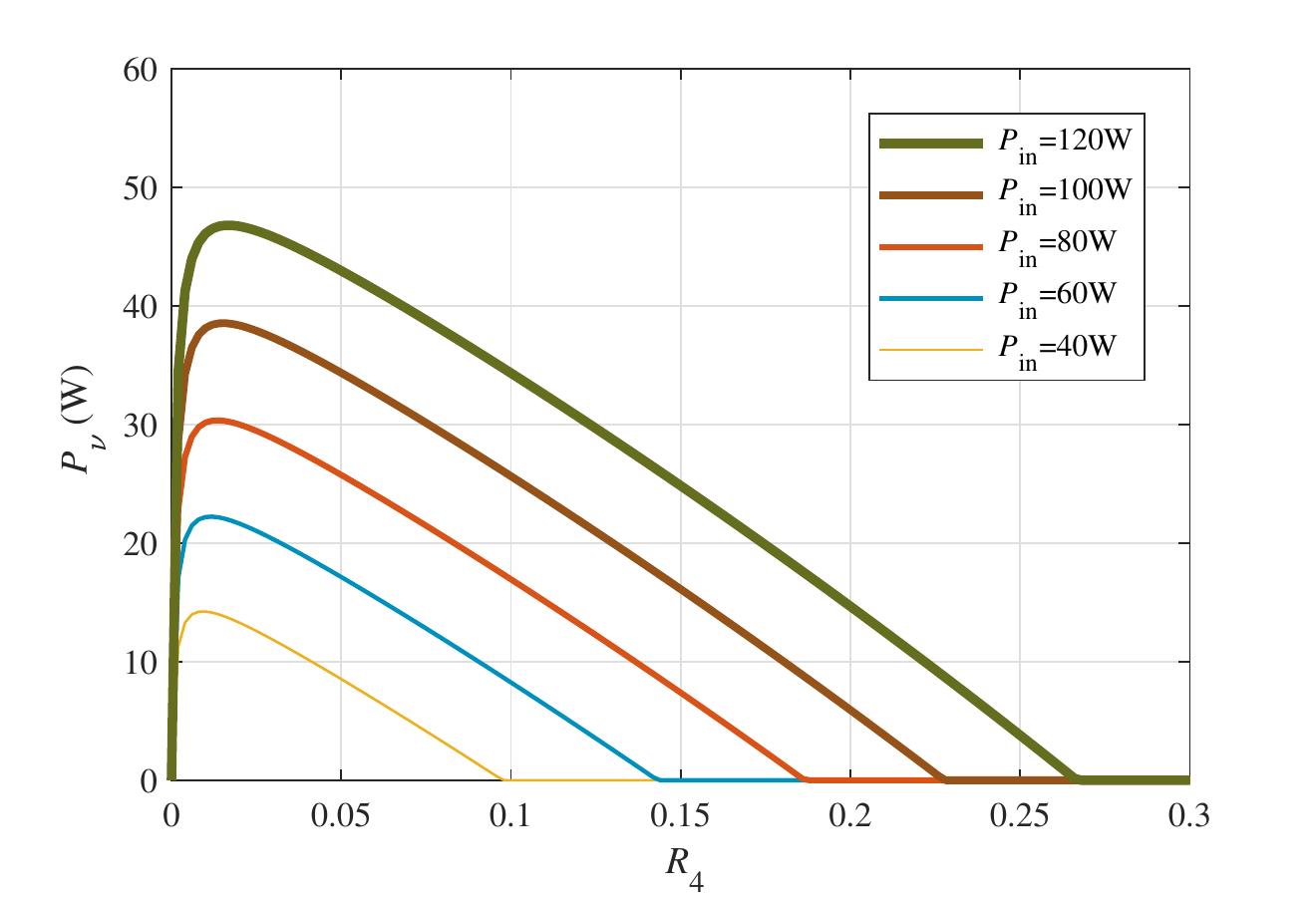}
	\caption{Fundamental beam power $P_{\nu}$ {\it vs}. splitter's reflectivity $R_4$ (transmission distance $d=5$ m, focal length of lens $f=0.03$~m, $f_{\rm RR}=3$~m)}
	\label{fig:Pnu-R4}
\end{figure}

\begin{table*} 
	\small
	\caption{Technology Comparison}
	\renewcommand{\arraystretch}{1}
	\centering
	\setlength{\tabcolsep}{1.5mm}
	\begin{tabular}{c c c c c c c c}
		\hline
		Work& Transmitter & Distance & Achievable Rate  & Beam Diameter &  Mobile Scheme & Positioning \\
		\hline
		Wang \textit{et al.}\cite{a201130.03}&wavelength swept laser&  1.4 m & 12 Gbit/s & - & tilted fiber grating & need\\ 
		Sung \textit{et al.}\cite{a201130.02}&XFP transceiver & 1.7 m  & 10 Gbit/s  & 114 mm  & fiber array & need\\ 
		Rhee \textit{et al.}\cite{a201201.02}&monochromatic laser diode&  3m & 32 Gbit/s  & 42 mm & optical phase array & need \\ 
		Chun \textit{et al.}~\cite{a201201.03}& four-color laser diode & $4$~m & $35$~Gbit/s & $10$~mm & MEMS mirror & need \\
		This work& RBCom system& 8 m & 11.2 Gbit/s & 6 mm & self-alignment & don't need\\ 
		\hline
	\end{tabular}
	\label{tab:comp}
\end{table*}

\subsection{Power of Fundamental Beam}
Figure \ref{fig:Pnu-R4} shows the power $P_\nu$ of the fundamental beam as a function of the {reflectivity $R_4$ of the coating M4}, for different electrical driving power $P_{\rm in}$ and a certain transmission distance $d=5$~m. As $R_4$ increases, $P_\nu$ increases rapidly at first and then decreases to $0$  gradually. The maximum $P_\nu$ can be obtained with a specific $R_4$. For instance, when $P_{\rm in}  = 100$~W, the maximum $P_\nu$ can be obtained when $R_4$ is close to $2$\%. Moreover, as $P_{\rm in}$ grows, the optimal value of $R_4$ appears slightly increasing.

\begin{figure} [t]
	\centering
	\includegraphics[width=3.4in]{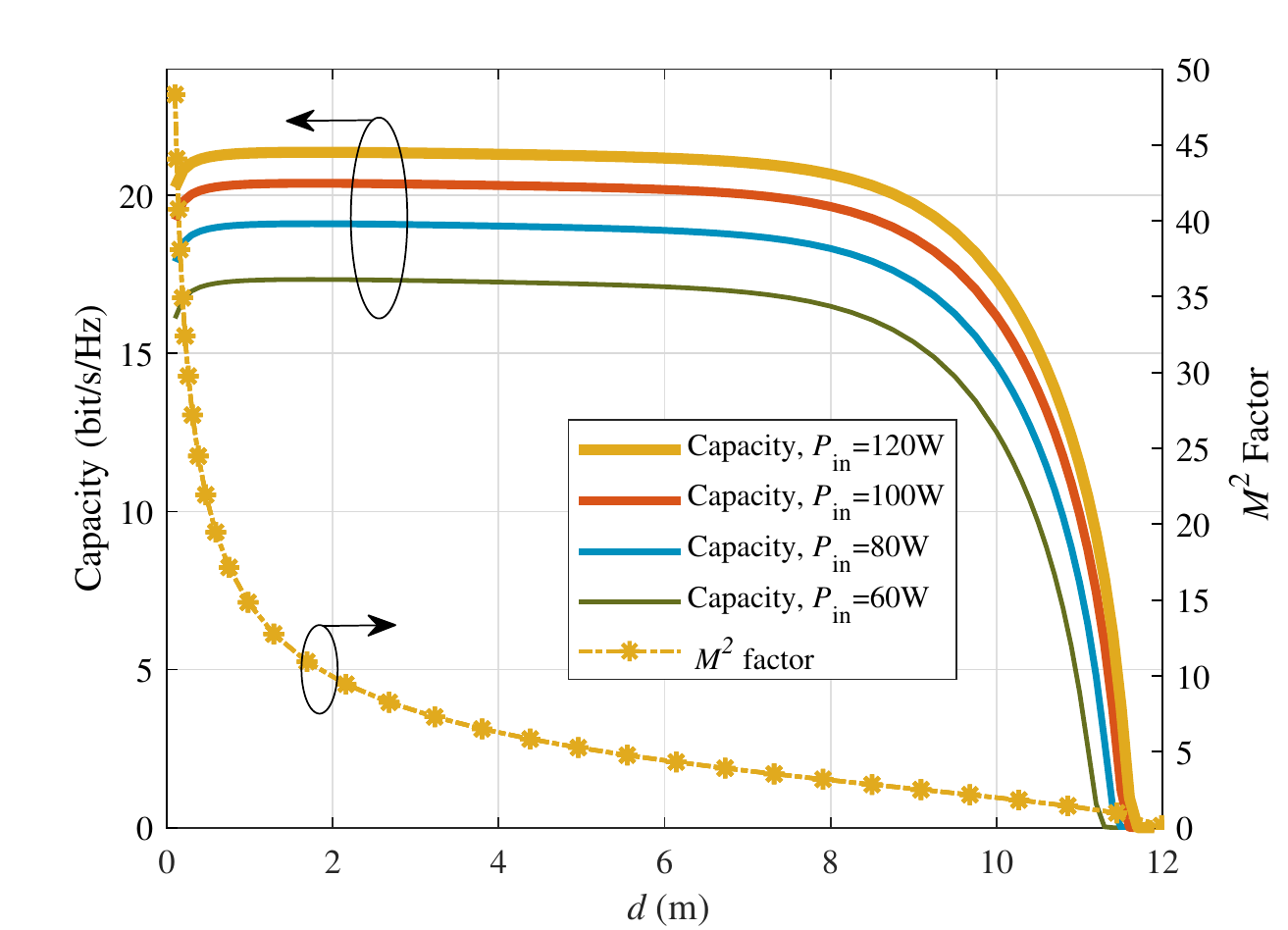}
	\caption{Channel capacity $\tilde{C}$ and $M^2$ factor {\it vs.} transmission distance $d$ (reflectivity $R_4=0.02$, focal length of lens $f=0.03$ m, $f_{\rm RR}=3$~m)}
	\label{fig:C-d}
\end{figure}

\subsection{Communication Performance}
We evaluated the communication capacity at different distance $d$. As depicted in Fig.~\ref{fig:C-d}, the channel capacity exhibits less difference for $d<8$~m. As $d$ grows from $8$~m to $12$~m, the capacity decreases to $0$ quickly. Furthermore, it is obvious that the $M^2$ factor keeps greater than $4$ for $d<6$~m, which indicates that at least $6$ transverse modes are oscillating together. In this case, the intra-cavity fundamental beam has a homogeneous transverse intensity distribution. When $d$ approaches $12$~m, the resonator turns to unstable state and the TEM$_{00}$ mode radius $w_{00}$ is greater than the gain medium radius (no matter how large the gain medium is). Therefore, the diffraction loss becomes so significant that no mode can keep oscillating. When the distance increases within $8$~m, the channel capacity decreases slightly.

\begin{figure}[t]
	\centering
	\includegraphics[width=3.4in]{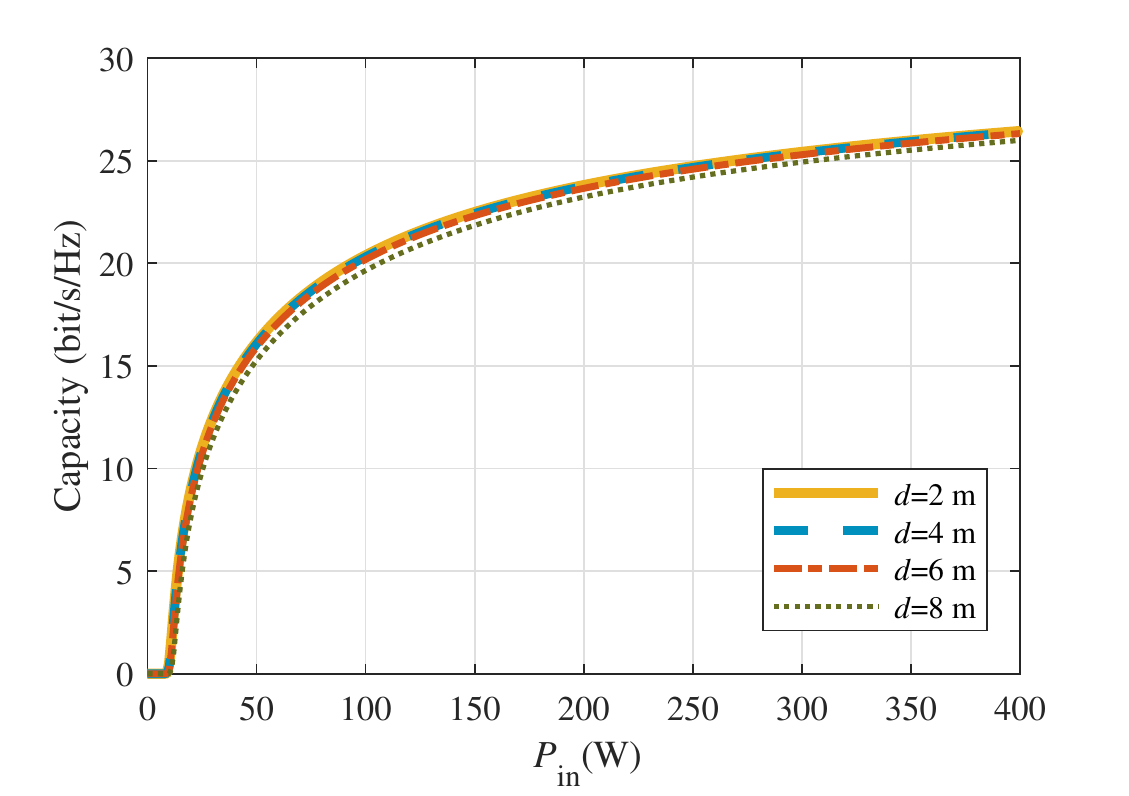}
	\caption{Channel capacity $\tilde{C}$ {\it vs.} electrical driving power $P_{\rm in}$ ($R_4=0.02$, focal length of lens $f=0.03$ m, $f_{\rm RR}=3$~m)}
	\label{fig:C-Psource}
\end{figure}

Figure \ref{fig:C-Psource} depicts the relationship between the channel capacity and the electrical driving power $P_{\rm in}$. There is a threshold of $P_{\rm in}$ for each case. If $P_{\rm in}$ is below the threshold, the resonance has not been established, and thus the capacity is $0$. After $P_{\rm in}$ exceeding the threshold, the channel capacity increases rapidly. Moreover, the growth rate of the capacity reduces gradually as $P_{\rm in}$ increases. When the electrical driving power is $100$~W and the distance is $8$~m, the channel capacity is $19.7$~bit/s/Hz. However, the channel capacity only increases to $26$~bit/s/Hz as the driving power reaches $400$~W. Within this range of driving power, the maximum detected power at the receiver is 35.4 mW. Compared with the fundamental beam power depicted in Fig.~\ref{fig:Pnu-R4}, the SHG efficiency is low. Even though, the achievable power level is enough for communication, while common LED-based VLC or laser communication systems generally receive less than $1$~mW due to beam divergence and/or safety issues~\cite{a201120.01}.

	In addition,  quadrature amplitude modulation (QAM)  and orthogonal frequency division multiplexing (OFDM) schemes are adopted in simulation to evaluate the communication performance. The QAM order is $16384$, which means each QAM symbol transmits $14$ bit. The OFDM signal contains $800$ subcarriers and is transmitted through $800$-MHz channel. The fast Fourier transform (FFT) length is $1024$. The cyclic prefix length is $176$. Besides, digital clipping is adopted to reduce the peak to average power ratio. Since the optical channel is a nonnegative real channel, the transmitted OFDM signal is biased by a direct-current (DC) signal.
	Result shows that when the pump source power is $120$~W and the transmission distance is $8$~m, the  data rate reaches 11.2 Gbit/s with bit error rate (BER) $<3.8\times10^{-3}$. The beam diameter at the receiver is approximate to $6$~mm. Table~\ref{tab:comp}  demonstrates the  distinction between our system and the state-of-art architectures.

\section{Discussion}
\label{sec:disc}

\subsection{Channel Bandwidth}
According to~\eqref{equ:ht}, the channel bandwidth is determined by the minimum bandwidth among which of the EOM, the air-transmission channel and the PD. For indoor application with intensity-modulation/direct-demodulation (IM/DD) scheme, the frequency characteristic of the air-transmission channel can be modeled as a direct-current (DC) gain. The multiple quantum well (MQW) EOM with bandwidth up to $37$~GHz was reported in~\cite{a200513.01}, which is higher than most of other EOMs. Generally, the modulating speed is limited by EOM's capacitance. Reducing the active area or increasing the intrinsic region thickness can decrease the capacitance. Thus, to produce a high-speed MQW EOM, we split the active area into small pixels. To reduce the power consumption, we can only select irradiated pixels for modulation operation, which benefits from the fact that the MQW can also perform as a PD if reversely biased~\cite{a190909.22}. Similar to the MQW EOM, the response speed of the PD is also determined by its diode's capacitance. For high speed communications with a large receiving aperture, PD arrays with small area are the best choice~\cite{a200430.04}, or one can adopt a lens to focus the light onto an independent small detector~\cite{a200513.11}. Commercially available PDs generally have bandwidth up to $20$~GHz~\cite{a200430.04}.

\subsection{Power Efficiency}
In order to improve the power efficiency, we can employ gain mediums with high saturated intensity and reduce the loss from the pumping progress. In this paper, we adopt an Nd:YVO$_4$ gain medium, as it has a high saturated intensity and is widely used in practice~\cite{a180720.02}. A better scheme is to employ semiconductor gain mediums which can be pumped directly with electricity so as to achieve a high efficiency~\cite{a200513.10}.
Besides, the gain medium's efficiency can be improved by reducing the gain medium radius; this way also improves the SHG efficiency, since a smaller gain medium cross section provides a smaller beam spot and a higher beam intensity. Although reducing the gain medium radius will increase the diffraction loss and decrease the overlap efficiency simultaneously, this issue can be addressed by adding optics to minify the TEM$_{00}$ mode radius in the gain medium. For example, one can design a coupled cavity with an internal focal lens~\cite{a200514.01}. To improve the SHG efficiency, the SHG medium length is supposed to be extended. Yet, this method also leads to a greater  retroreflector length, so it can be implemented for some applications without the needs of small assembly size.

\subsection{Carrier Wavelength}
Most long-range FSO systems operate in the wavelength windows of $780-850$~nm and $1520-1600$~nm, in order to ensure attenuation $<0.2$~dB/km~\cite{a200513.05}. These windows are nicely suitable to our system, as the former lies in the second-harmonic band of the latter. Although $532$-nm carrier was demonstrated in our system, most of the preferable wavelengths can be obtained by replacing the gain material, especially, with semiconductor devices~\cite{a180720.02, a200430.04}.

\subsection{Resonance Stability}
The fluctuation origins from two aspects: 1) Some  device parameters change while moving; for example, the reflectivity of mirrors varies with the incident angle, which induces slow power fluctuation of the intra-cavity beam; and 2) the balance condition of the resonance system is broken so that the oscillation turns into unstable state. In terms of the first aspect, the moving speed of the devices is quite slow, compared with the signal's swinging speed. This effect can be overcome easily by many common schemes adopted by the existing communication system, for example, automatic gain control (AGC)~\cite{a201201.04}. In terms of the second aspect, due to the open-cavity structure, moving and atmospheric turbulence can break the balance condition of the resonance, which leads to intensity fluctuation of the fundamental beam. This phenomenon is called relaxation oscillation in literature. The oscillation frequency and the dumping time depend on the pumping rate, the upper-level atoms lifetime, and the cavity photons lifetime~\cite{a181224.01}. Slow-frequency and small-amplitude  oscillation can be achieved by supplying an adequate electrical driving power that is just above the threshold. Some signal  processing and coding schemes can  also be introduced to overcome the  impacts brought from relaxation  oscillation. Besides, a more  effective way to suppress fluctuation is adopting optoelectronic feedback controlling~\cite{a200514.02},  especially for solid-state laser which usually fluctuates at a few kHz~\cite{a200515.01}. This scheme does not need  channel information estimation, but only needs to detect the intra-cavity beam power at the transmitter and correspondingly control the pump source power.

\subsection{Uplink Communication}
Uplink communication can be realized by improving the system design. For example, an extra SHG medium is mounted at the receiver with correct phase matching angle. Thus, the forth harmonic beam can be acquired, which acts as the uplink communication carrier. Or we can design a receiver identical to the transmitter, excluding the gain medium, to generate a second harmonic beam. Moreover, we can mount liquid shutters between the real mirror and the EOM at both the transmitter and receiver to switch the beam propagating path. Uplink communication  can be realized by closing the transmitter's shutter and opening the receiver's shutter, which only allows the second harmonic beam to propagate from the receiver to the transmitter.

\section{Conclusions}
\label{sec:con}
In this paper, we aimed at dealing with the echo-interference issue in the RBCom system. We at first proposed an echo-interference-free RBCom system design which exploits SHG to produce frequency-doubled carrier beam for communications. Next, we established  an analytical model of the proposed system. Moreover, we evaluated the beam radius in the resonator and the channel capacity of the RBCom system. Communication simulation with OFDM scheme was conducted. The results show that our system achieves longer transmission distance and smaller beam diameter for  transmission beyond $10$ Gbit/s, compared with the existing OWC technologies. Besides, our system avoids the positioning progress with the aid of its self-alignment feature. Further studies on the mobility and stability and the approach of improving the power efficiency are well motivated.

\appendices
\section{Diffraction Loss Approximation}
\label{sec:diffLossAppr}

\begin{figure}
	\centering
	\includegraphics[width=2.8in]{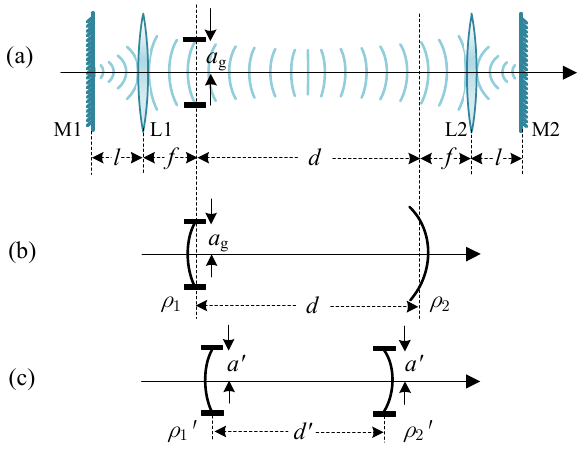}
	\caption{Equivalent resonator for diffraction loss calculation }
	\label{fig:equReso}
\end{figure}

Obtaining diffraction loss of the spatially separated laser resonator which consists of several elements and an internal aperture is complicated generally through numerical method. Here we provide a method to obtain an approximate solution. Note that all the diffraction loss are assumed to be caused by the gain medium, as its aperture is much smaller than that of other lenses and mirrors in the system. 

As depicted in Fig.~\ref{fig:equReso}(a), the total diffraction loss for a round-trip is composed of the diffraction losses on the two sides of the aperture. From~\eqref{equ:MRR}, the retroreflector placed on the left side of the gain medium  has a vanished B-element in its ray transfer matrix. For this kind of system, its diffraction loss can be neglected~\cite{a200511.01}.

Then we create an equivalent empty resonator to obtain the diffraction loss on the right surface of the gain medium, as shown in Fig.~\ref{fig:equReso}(b). The mirrors surface of the equivalent  resonator is assumed identical to the constant-phase surface of the original intra-cavity beam at two selected locations, i.e., the location of the gain medium (located at the pupil of the retroreflector RR1) and the pupil of the retroreflector RR2. Thus, the length of the equivalent resonator is $d$.

According to the formula for calculating the ROC of the constant-phase surface of a Gaussian beam, the ROCs of the mirrors of the equivalent resonator is obtained as~\cite{a181221.01}
\begin{equation}
	\rho_1=-\dfrac{1}{\Re\left[1/q(l+f)\right]},~~~
	\rho_2=\dfrac{1}{\Re\left[1/q(l+f+d)\right]},
\end{equation}
where $\Re[\cdot]$ takes the imaginary part of a complex quantity.
The resonator with one finite-aperture mirror and one infinite-aperture mirror is equivalent to another symmetric resonator with two identical apertures at both mirrors, as depicted in Fig.~\ref{fig:equReso}(c). The equivalent Fresnel number $N'$ and g-parameters \{$g_1'$, $g_2'$\} are~\cite{a200224.01}
\begin{align}
	N'&=\dfrac{a_g^2}{2\lambda d(1-\dfrac{d}{\rho_2})}, \nonumber\\
g_1'&=g_2'=1-2d\left(\dfrac{1}{\rho_1}+\dfrac{1}{\rho_2} -\dfrac{d}{\rho_1\rho_2}  \right).
\end{align}
Note that the Fresnel number taking a negative value is acceptable, as it only indicates the change of the phase rather than the intensity pattern~\cite{a200519.01}. Then, the round-trip diffraction loss factor of the TEM$_{00}$ mode is approximated by~\cite{a190318.99}
\begin{equation}
\Gamma_{\rm diff}=1-\exp\left[-2\pi \left|N'\right|~\sqrt{\dfrac{g_1'(1-g_1'g_2')}{g_2'}}\right].
\end{equation}
In this paper, we only analyze the diffraction loss of the TEM$_{00}$ mode, as it determines the threshold of the resonance. In fact, for the operations with multiple transverse modes, the diffraction loss is small and negligible.



%

%



\ifCLASSOPTIONcaptionsoff
  \newpage
\fi




\bibliographystyle{IEEETran}
\small
%
\bibliography{mybib}
%
%

%

%
%







\end{document}